\documentclass[twocolumn,english,prb]{revtex4-1}

\usepackage{textcomp}
\usepackage{amsmath}
\usepackage{graphicx, subfigure}
\usepackage{amssymb}
\usepackage{amsmath}
\usepackage{amsbsy}
\usepackage{eucal}
\usepackage{relsize}
\usepackage{bm}
\usepackage{setspace}
\usepackage[usenames,dvipsnames]{color}
\usepackage{mathptmx}

\usepackage{color}

\newcommand{\bsh}{{\mathbf{S}}}
\newcommand{\sh}{{S}}

\newcommand{\Jfm}{J_{\mathrm{core}}}
\newcommand{\Jafm}{J_{\mathrm{edge}}}
\newcommand{\Jint}{J_{\mathrm{ce}}}
\newcommand{\Kfm}{k_{\mathrm{core}}}
\newcommand{\Kafm}{k_{\mathrm{edge}}}

\newcommand{\bHapp}{\mathbf{H}_{\mathrm{app}}}

\begin{document}

\title{Atomistic modeling of magnetization reversal modes in $L1_{0}$ FePt nanodots with magnetically soft edges}
\author{Jung-Wei Liao$^{1,2}$}
\author{Unai Atxitia$^{2,3}$}\altaffiliation[Present address: ]{Department of Physics, University of Konstanz, Konstanz D-78457, Germany}
\author{Richard F. L. Evans$^{2}$}\email{richard.evans@york.ac.uk}
\author{Roy W. Chantrell$^2$}
\author{Chih-Huang Lai$^1$}\email{chlai@mx.nthu.edu.tw}
\affiliation{$^1$Department of Materials Science and Engineering, National Tsing Hua University, Hsinchu 30013, Taiwan}
\affiliation{$^2$Department of Physics, The University of York, York YO10 5DD, United Kingdom}
\affiliation{$^3$Departamento de Fisica de Materiales, Universidad del Pais Vasco, UPV/EHU, 20018 San Sebastian, Spain}
\date{\today}

\begin{abstract}
Nanopatterned FePt nano-dots often exhibit low coercivity and a broad switching field distribution, which could arise due to edge damage during the patterning process causing a reduction in the $L1_{0}$ ordering required for a high magnetocrystalline anisotropy. Using an atomistic spin model, we study the magnetization reversal behavior of $L1_{0}$ FePt nanodots with soft magnetic edges. We show that reversal is initiated by nucleation for the whole range of edge widths studied. For narrow soft edges the individual nucleation events dominate reversal; for wider edges, multiple nucleation at the edge creates a circular domain wall at the interface which precedes complete reversal. Our simulations compare well with available analytical theories. The increased edge width further reduces and saturates the required nucleation field. The nucleation field and the activation volume manipulate the thermally induced switching field distribution. By control of the properties of dot edges using proper patterning methods, it should be possible to realize exchange spring bit patterned media without additional soft layers.
\end{abstract}

\pacs{}
\keywords{}

\maketitle

\section{Introduction}
Continuing requirements for greater digital data storage capacity has lead to continued growth in data storage density in magnetic recording media. Future improvements are limited by the magnetic recording trilemma, caused by competing requirements of reduced signal to noise ratio, thermal stability of written information, and writability.\cite{Weller-IEEE-1999} Two solutions for the magnetic recording trilemma have been proposed: first heat assisted magnetic recording (HAMR),\cite{Rottmayer-IEEE-2006, McDaniel-2005} where laser heating during recording is used to lower the anisotropy sufficiently to achieve writing; the second solution is bit-patterned media (BPM), where each bit is defined by a single dot in a lithographically defined array\cite{IEEE-Trans-Magn-33-990-1997} and the larger magnetic volume reduces the requirement for high anisotropy required for long-term thermal stability. 

Bit patterned media can be made using a variety of methods including patterning\cite{Science-280-1919-1998, PRB-78-024414-2008, APL-99-062505-2011, APL-101-092402-2012, APL-100-102401-2012, JMMM-324-3737-2012, IEEE-49-693-2013} and self-assembly of magnetic nanoparticles.\cite{Sun17032000} Controlling the microstructural properties of magnetic nanoparticles is quite challenging, however lithographic patterning techniques allow a continuous $L1_{0}$ FePt film to be patterned into an array of isolated magnetic islands or 'dots'.\cite{Science-280-1919-1998, PRB-78-024414-2008, APL-99-062505-2011, APL-101-092402-2012, APL-100-102401-2012, JMMM-324-3737-2012, IEEE-49-693-2013} However, during lithography ions near the dot edge can reduce the $L1_{0}$ ordering, resulting in magnetically soft edges.\cite{PRB-78-024414-2008, APL-99-062505-2011, JMMM-324-3737-2012, APL-100-102401-2012, IEEE-49-693-2013} The presence of damaged edges in the dots could reduce both the coercivity\cite{PRB-78-024414-2008, JMMM-324-3737-2012, IEEE-49-693-2013} and the thermal stability.\cite{APL-100-102401-2012}

In addition to decreasing the coercivity, a broad switching field distribution (SFD) can also lead to write errors in neighboring bits during the writing process. The SFD (the variation of switching fields between dots) includes both extrinsic and intrinsic components.\cite{APL-90-162516-2007, PhysRevLett-96-257204-2007, Nanotech-21-035703-2010, JAP-112-023903-2012, APL-101-182405-2012} The extrinsic SFD may be caused by dipolar interaction between dots, and the intrinsic SFD arises from variations of intrinsic magnetic properties of dots, including anisotropy $K$, volume $V$, and the easy axis alignment.\cite{PhysRevLett-96-257204-2007} Furthermore, thermal fluctuations also broaden the intrinsic SFD, known as the thermal SFD.\cite{Nanotech-21-035703-2010, JAP-112-023903-2012, APL-101-182405-2012}
Within the simple Stoner-Wohlfarth approximation (monodomain), the thermal SFD is mainly related to the anisotropy energy barrier, $KV$, and the measurement time scale. This makes the thermal SFD pronounced at high field sweep rates associated with the recording process.\cite{JAP-112-023903-2012, APL-101-182405-2012}

Reversal behavior in relatively large dots with magnetically soft edges of fixed width, associated with ion-damage from the etching process, has been studied previously,\cite{PhysRevB-84-214427-2011} where the presence of soft edges was shown to change the reversal mode. In addition, small-sized dots with ring-shaped soft edges of varied width have been investigated by macrospin analytic models without including the thermal fluctuations.\cite{Vokoun-2010} The increased width of edge is found to reduce the coercivity of dots,\cite{Vokoun-2010} suggesting a strong relationship between the edge width and the reversal mode. Further control of the magnetic properties of edges with fixed width in large sized dots can also be done via soft He$^{+}$ irradiation.\cite{PhysRevB-85-214417-2012} The experimental observations can only be explained by the model including the thermal fluctuations.\cite{PhysRevB-85-214417-2012} All the reported works indicate that either the edge width\cite{Vokoun-2010} or the thermal fluctuations\cite{PhysRevB-85-214417-2012} affects the reversal mode and could result in different switching field distributions of patterned dots. However, the edge-width dependence of the reversal mode including thermal fluctuations is still not understood.

Here we develop a computational model to study magnetization reversal modes in $L1_{0}$ FePt dots with magnetically soft edges. We employ an atomistic spin model formalism, which provides detailed information on reversal modes unreachable by standard micromagnetic simulations.\cite{APL-87-122501-2005, barker:192504} In particular, soft edges of only a few nanometers are tractable, and we can further study the effect of the reduced exchange coupling at the interface, possibly resulting from the core/edge interface roughness. Moreover, thermal effects are consistently taken into account within our model using the Langevin dynamics formalism that allows us to study the relationship between the coercivity, the thermal SFD, and reversal modes. We further compare the atomistic spin modeled results with available analytic approaches, which were originally presented for hard/soft nano wires,\cite{goto:2951, PhysRevB.70.104405, Suess2009545, Physica-B-319-122-2002, APL-89-113105-2006, APL-89-062512-2006} to examine the validity of these approaches for the core/shell nanostructure. These simpler approaches are capable of highlighting the key physics. Additionally, since the atomistic resolution in the simulation makes this method computationally intensive restricting the size of the calculated system to nanometer length scales, these validated analytic approaches could be potentially utilized to investigate properties of large sized systems, e.g., a dot array.
	
\section{Atomistic spin model}
\begin{figure}[!t]
\center
\includegraphics[width=8cm]{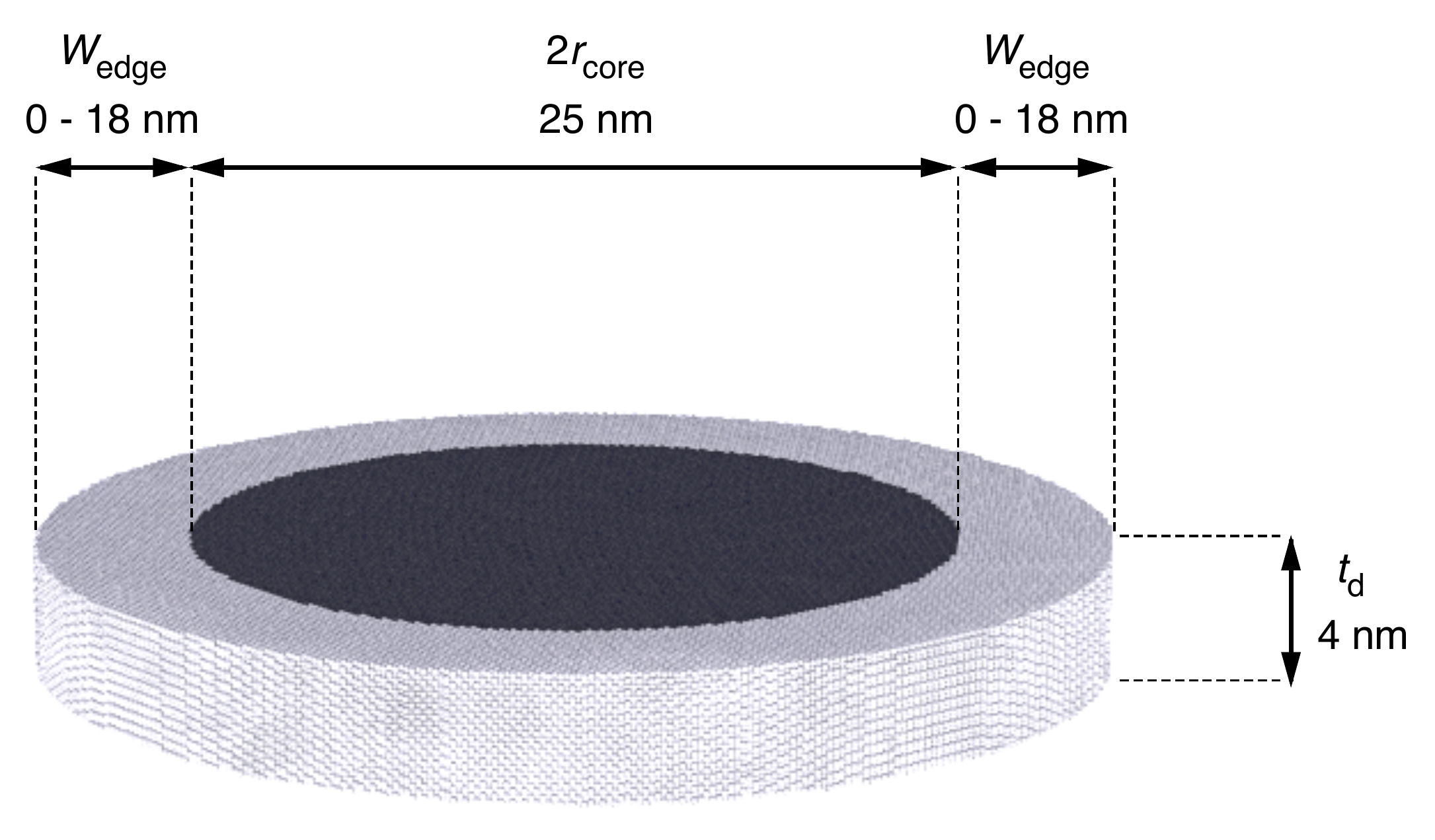}
\caption{(Color online) Schematic diagram of the atomistic modeled dot. Dark and white gray regions represent the core and the edge atoms, respectively.}
\label{fig:Simulated-dot}
\end{figure}

The studied nanodots are composed of a magnetically hard core and a magnetically soft edge, as illustrated in Fig.~\ref{fig:Simulated-dot}. In the case of patterned dots, we hypothesize that the edge region loses its $L1_0$ atomic order due to the patterning process, making it magnetically soft. We focus exclusively on the width of the edge $W_\mathrm{edge}$, with the fixed core size, $r_\mathrm{core}$, on the magnetization reversal. We therefore fix the diameter of the core $2r_\mathrm{core} = 25$ nm and the dot thickness, $t_\mathrm{d}=4$ nm, while the edge width is varied systematically from $W_\mathrm{edge} = 0-18$ nm. We note that in this approach the different $W_\mathrm{edge}$ varies the total volume of dot and therefore affects the corresponding thermal stability, which is beyond the scope of the present work. The system is constructed from a single face-centered cubic crystal and cut into the shape of a nanodot with the desired geometry.

The nano dots are modeled using an atomistic spin model approach\cite{atomistic-spin-model} with the \textsc{vampire} software package.\cite{vampire-url} The energetics of the system are described by the spin Hamiltonian with the Heisenberg exchange, given by:
%=============================================
\begin{eqnarray}
\label{HamiltonianA}
\mathcal{H} &=& \mathcal{H}_{\mathrm{core}} + \mathcal{H}_{\mathrm{edge}} \\
\nonumber
\mathcal{H}_{\mathrm{core}} &=&  -\sum_{i,j} \Jfm \bsh_i \cdot \bsh_j - \sum_{i,\nu} \Jint\bsh_i \cdot \bsh_{\nu} - \\
\label{HamiltonianB}
 && \Kfm \sum_i \left(\sh_i^z\right)^2 - \mu_{\mathrm{core}} \sum_i  \bHapp \cdot \bsh_i \\
\nonumber
\mathcal{H}_{\mathrm{edge}} &=& -\sum_{\nu,\delta} \Jafm \bsh_{\nu} \cdot \bsh_{\delta} - \sum_{\nu,j} \Jint \bsh_{\nu} \cdot \bsh_{j} - \\
\label{HamiltonianC}
 &&  \Kafm \sum_{\nu} \left(\sh_{\nu}^{z}\right)^2 - \mu_{\mathrm{edge}} \sum_{\nu} \bHapp \cdot \bsh_{\nu}
\end{eqnarray}
%==============================================
where $ \bsh = \boldsymbol{\mu}/\mu$ are spin unit vectors, $i,j$ label core sites with moment $\mu_{\mathrm{core}}$,  and $\nu,\delta$ label edge sites with moment $\mu_{\mathrm{edge}}$. Here we assume the same moment for both core and edge such that $\mu_{\mathrm{core}} = \mu_{\mathrm{edge}} = 1.5 \; \mu_{\mathrm{B}}$, which compares well to the saturation magnetization of $L1_0$ FePt  as obtained in  experiment.\cite{PhysRevB.66.024413} $\Jfm$ and $\Jafm$ are the exchange interactions between moments of the same type in the core and the edge, respectively. We consider
 only nearest neighbor interactions between the moments. We select values of the exchange energy  to give a Curie temperature around 700 K comparable with experiment, namely $\Jfm = \Jafm = 3\times10^{-21}$ J/link. $J_\mathrm{ce}$ represents the interfacial exchange interaction between the core and the edge and is varied as a parameter between 0 and $\Jfm$. $\Kfm = 4.9\times10^{-23}$ J/atom is the uniaxial anisotropy constant of the core spins (with easy axis perpendicular to the film plane) and $\Kafm = 1 \times10^{-24}$ J/atom is the uniaxial anisotropy of the edge spins. $\bHapp$ is the external applied field.

%=============================================
%=============================================

The hysteresis loops are calculated dynamically using the stochastic Landau-Lifshitz-Gilbert (LLG) equation at the atomic level, given by
%=========================================
\begin{eqnarray}
\label{eqn:LLG}
	\frac{\partial\bold{S}_i}{\partial t} = -\frac{\gamma}{(1+\lambda^{2})} \bold{S}_i \times \big[\bold{H}_{i, \mathrm{eff}}+ \lambda (\bold{S}_i \times \bold{H}_{i, \mathrm{eff}})\big]\mathrm{,}
\end{eqnarray}
%=========================================
where $\lambda$ is the intrinsic damping parameter, $\gamma = 1.76 \times 10^{11}$ T$^{-1}$s$^{-1}$ is the absolute value of the gyromagnetic ratio, and $\bold{H}_{i, \mathrm{eff}}$ is the effective magnetic field in each spin. The field is derived from the spin Hamiltonian and is given by
%=========================================
\begin{equation}\label{eqn:Heffth}
  \bold{H}_{i, \mathrm{eff}} = -\frac{1}{\mu_i}\frac{\partial \mathcal{H}}{\partial \bold{S}_i}
  + \bold{H}_{\mathrm{demag}, i} + \bold{H}_{i, \mathrm{th}}\mathrm{,}
\end{equation}
%=========================================
where $\bold{H}_{\mathrm{demag}, i}$ and $\bold{H}_{i, \mathrm{th}}$ are the demagnetization and the thermal fields, respectively.
Since the calculation of the demagnetization field at the atomic level is computationally expensive, we have instead calculated the demagnetization field by applying the approach developed by Boerner \textit{et al}.\cite{IEEE-Trans-Magn-41-936-2005} Within this approach, the dot is divided into regular macrocells with the volume $V_k = (1.77)^{3}$ nm$^{3}$ which contains $250$ atomic spins. The value of spin's moments within each macrocell are then summed to obtain the macrocell magnetic moment, $\boldsymbol{\mu}_k = \sum_{\delta \in \triangle _{k}}\mu_{\delta}\bold{S}_{\delta}$, where $k$ labels macrocell sites, and $\delta$ labels spin sites in each macrocell, $\triangle _{k}$. We then calculate the demagnetization field of each macrocell, $\bold{H}_{\mathrm{demag}, k}$, by using the corresponding magnetic moment and treat it as the demagnetization field of each spin in the macrocell, $\bold{H}_{\mathrm{demag}, i}$. $\bold{H}_{\mathrm{demag}, k}$ is calculated by direct pairwise summation including the macrocell self-demagnetization\cite{atomistic-spin-model}

%=============================================
\begin{equation}\label{Demag-field}
	\bold{H}_{\mathrm{demag}, k} = \frac{\mu_0}{4\pi}\sum_{k \neq l}\frac{3(\boldsymbol{\mu}_l \cdot \boldsymbol{\hat{r}}_{kl})\boldsymbol{\hat{r}}_{kl} - \boldsymbol{\mu}_l}{|\boldsymbol{r}_{kl}|^{3}} - \frac{\mu_0}{3}\frac{\boldsymbol{\mu}_k}{V_k}
\end{equation}
%=============================================
where $\mu_0=4\pi \times 10^{-7}$ T$^2$J$^{-1}$m$^3$ is the vacuum permeability, $\boldsymbol{r}_{kl}$ is the vector between $k$ and $l$ macrocell sites, and $\boldsymbol{\hat{r}}_{kl}=\boldsymbol{r}_{kl}/|\boldsymbol{r}_{kl}|$ is the corresponding unit vector. This is a computationally efficient approach since the number of macrocells is relatively small and moreover, since the magnetostatic field varies rather slowly with time it needs updating only on a timescale of around 1000 time steps.\cite{atomistic-spin-model}
 The thermal fluctuations are represented using Langevin dynamics, \cite{PhysRev.130.1677, Langevin-dynamics} where the thermal field $\bold{H}_{i, \mathrm{th}}$ is given by
%==============================================
\begin{equation}\label{thermal-field}
\bold{H}_{i, \mathrm{th}} = \boldsymbol{\Gamma}(t) \sqrt{\frac{2 \lambda k_{\mathrm{B}}T}{\gamma \mu_i \Delta t}}\mathrm{,}
\end{equation}
%==============================================
where  $k_\mathrm{B}$ is the Boltzmann constant, $T$ is the heat bath
temperature, $\lambda$ is the Gilbert damping parameter, $\gamma$
is the absolute value of the gyromagnetic ratio, and $\Delta t$ is the integration time step. The thermal fluctuations are represented by a vector Gaussian distribution in space $\boldsymbol{\Gamma}(t)$ with a mean of zero and generated from a pseudo-random number generator. The simulations in this work are carried out at a heat-bath temperature of $T=300$ K. We set the damping parameter $\lambda = 1.0$ to reduce the computational time required for reaching an equilibrium state. The LLG equation is integrated using the Heun integration scheme\cite{Langevin-dynamics} with an integration time step $\Delta t=1$ fs.

%============================================

\section{Results}

%============================================

%==============================================
\begin{figure}[!t]
\center
\includegraphics[width=8cm]{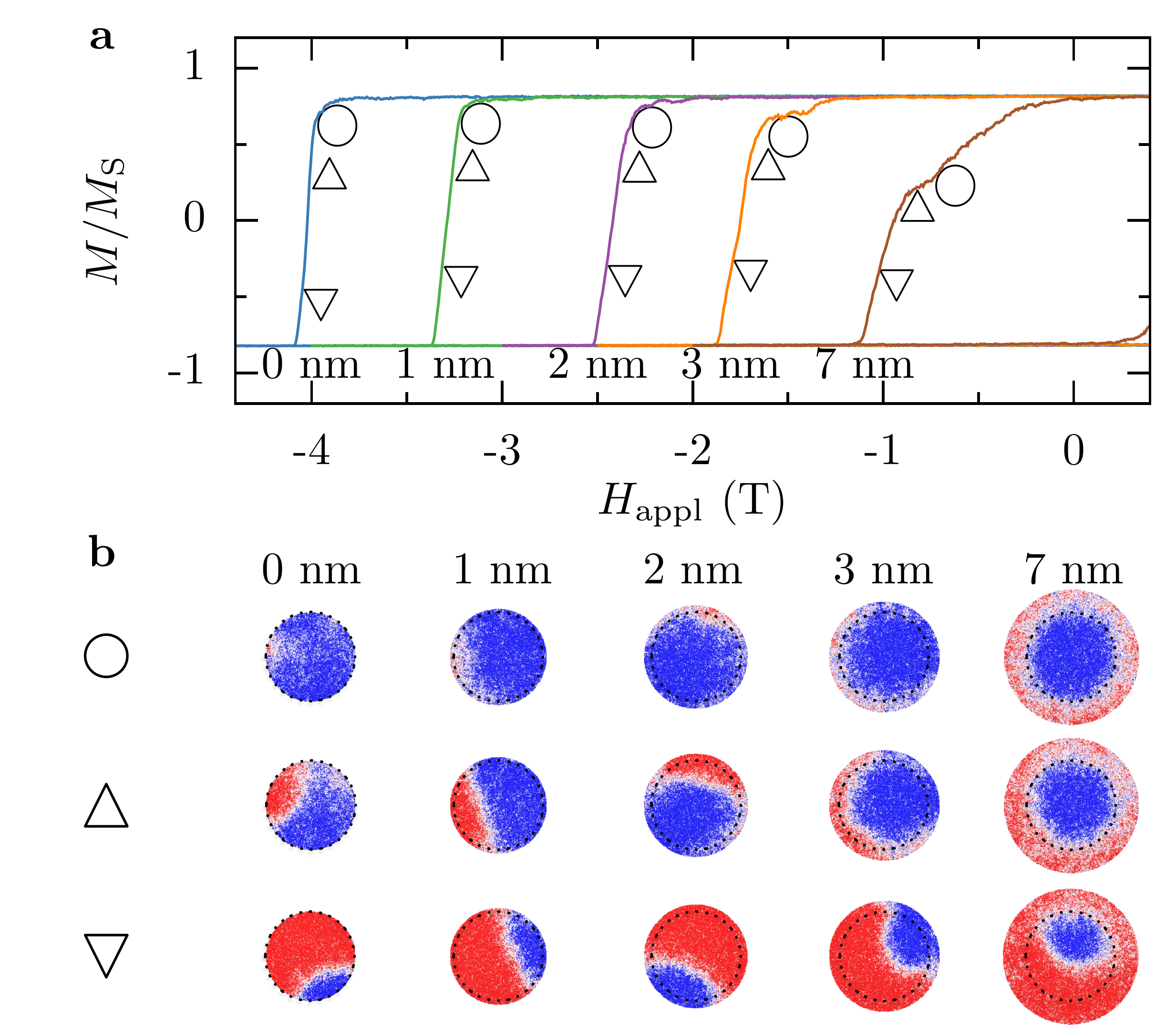}	
\caption{(Color online) (a) Simulated out-of-plane hysteresis loops for dots with different edge widths. Magnetization is normalized to the saturation magnetization at $0$ K. Snapshots of domain configurations during reversal, observed along the dot plane normal direction, are shown in (b). Symbols in (a) and on the left in (b) indicate the position of snapshots during the reversal process. The color scale (blue to red) represents the magnetization component along the easy axis direction. Black dotted circles denote the position of the core/edge interface.}
\label{fig:properties}
\end{figure}
%==============================================

In order to study reversal modes we simulate hysteresis loops  as a function of the width of edges, $W_\mathrm{edge}$. To calculate the hysteresis loops we apply an external field in a range from $-5$ to $+5$ T, which lies above the anisotropy field in the core, at intervals of $5$ mT. The field sweep rate is 5 T/ns.
Initially we consider that the interfacial core and edge spins are strongly coupled by setting $J_\mathrm{ce}=J_\mathrm{core}=J_\mathrm{edge}=3.0\times10^{-21}$ J/link.

\begin{figure}[!t]
\center
\includegraphics[width=8cm]{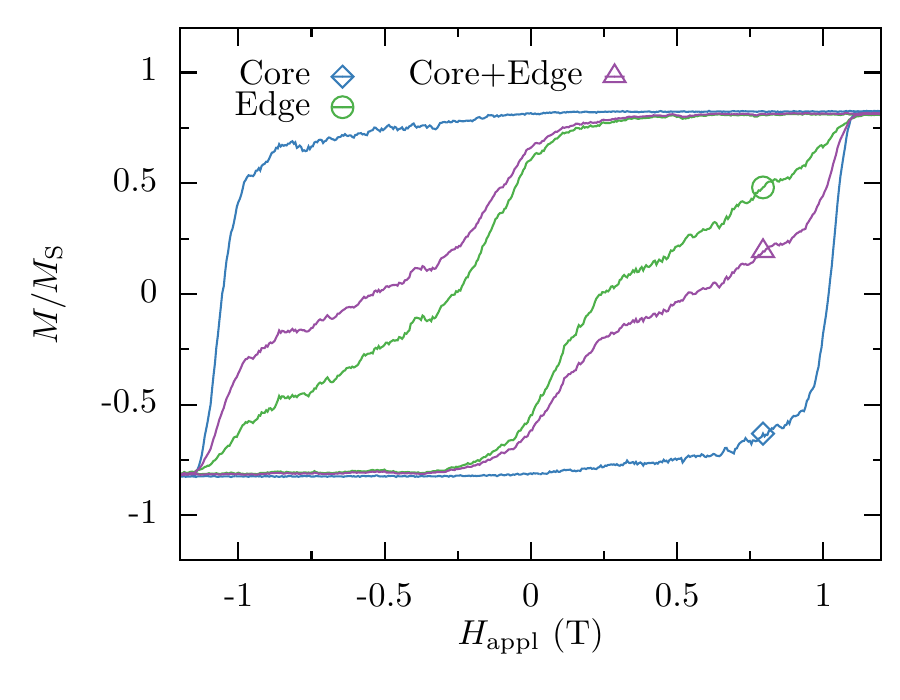}	
\caption{ (Color online) Whole hysteresis loop of the dot with $W_\mathrm{edge}=12$ nm and two individual loops of its core and its edge, decomposing the whole loop.}
\label{fig:individual-hysteresis}
\end{figure}

Figure~\ref{fig:properties}(a) shows representative out-of-plane hysteresis loops for a range of edge widths. One can observe that by increasing edge width, both the nucleation and coercive fields decrease. Furthermore, the square-like hysteresis loop for narrow edges turns into a two-step reversal as the edge width increases, indicating a change in the reversal process. Figure~\ref{fig:properties}(b) illustrates the corresponding snapshots of spin configurations during reversal for various $W_\mathrm{edge}$. The reversal mode strongly depends on the edge width, which will be discussed in more detail in the following sections. To obtain detailed information on the observed reversal behavior, we also calculate hysteresis loops for  each edge width for 30 different realizations of the random number generator. Therefore, we average over 60 statistically independent values to obtain the mean coercivity and standard deviation. Since we are considering dots with the same magnetic properties in our simulations, the deviation from the mean arises completely from the thermal fluctuations. Thus the standard deviation is a manifestation of the intrinsic SFD resulting from thermal fluctuations.~\cite{APL-101-182405-2012} This is an important parameter since it increases with increasing field sweep rate and is significant at timescales associated with data transfer in information storage.
Additionally, we  separately calculate the coercivity fields of both the core, $H_\mathrm{c}^\mathrm{core}$, and the edge, $H_\mathrm{c}^\mathrm{edge}$, shown in Fig.~\ref{fig:individual-hysteresis}.

%==============================================
%==============================================

To do so, we  calculate the individual reduced magnetization of the core and edge as follows,
%====================================================
\begin{equation}
\boldsymbol{\mu}_\mathrm{core}=\frac{|\mu_\mathrm{core}|} {N_\mathrm{core}}
\sum_{i \in \mathrm{core}}\bold{S}_i, \quad
\boldsymbol{\mu}_\mathrm{edge}=\frac{|\mu_\mathrm{edge}|}{N_\mathrm{edge}}\sum_{i \in \mathrm{edge}}\bold{S}_i ,
\end{equation}
%======================================================
where $N_\mathrm{core(edge)}$ denotes the number of atoms in the core (edge).

%==============================================
\begin{figure}[!t]
\center
\includegraphics[width=8cm]{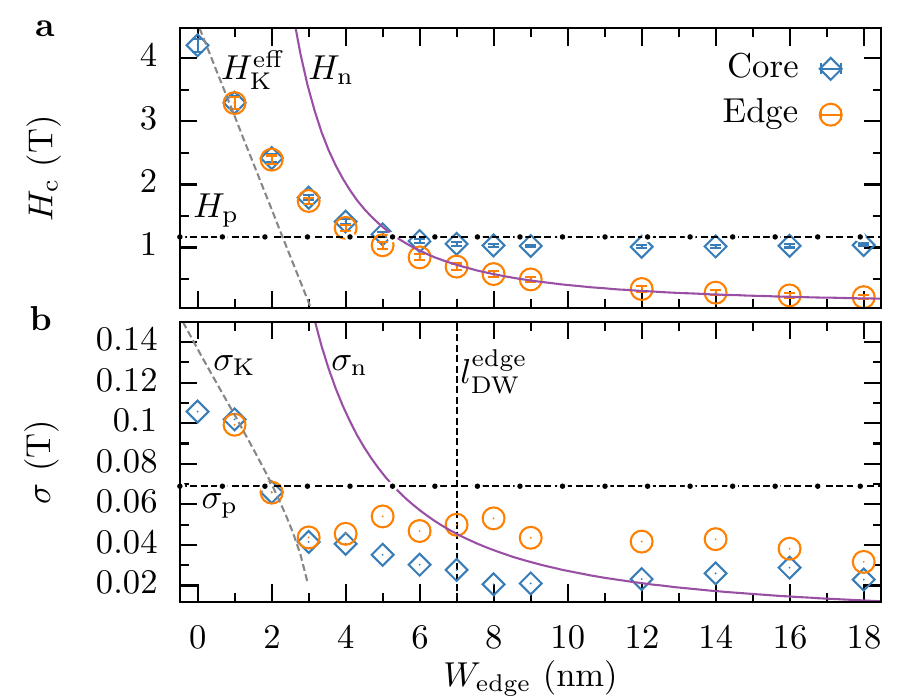}	
\caption{ (Color online) (a) Mean coercivity of both the core and the edge as a function of the width of edges. The gray dashed line represents the effective anisotropy field calculated by the linear chain model. Purple solid and black dashed lines indicate the nucleation and pinning fields respectively. (b) Standard deviation of coercivity of both the core and the edge as a function of the width of edges giving an estimate of the thermal switching field distribution. The gray dashed line represents the deviation approached by the effective anisotropy field. Purple solid and black dashed lines are the deviation calculated by the nucleation and the pinning fields. The vertical dashed line denotes the domain-wall length in the edge.}
\label{fig:coercivity-deviation}
\end{figure}
%==============================================

Figure~\ref{fig:coercivity-deviation}(a) shows the variation of the mean coercivity $H_\mathrm{c}^\mathrm{core(edge)}$ as a function of $W_\mathrm{edge}$, and the corresponding standard deviation, $\sigma_\mathrm{core(edge)}$, is shown in Fig.~\ref{fig:coercivity-deviation}(b). The coercivities and the standard deviation are strongly dependent on the edge width, which will be discussed in the following sections.
Furthermore, to understand the reversal mode, we will compare coercive fields obtained from atomistic spin model simulations with those obtained from different theoretical approaches,\cite{goto:2951, PhysRevB.70.104405, Suess2009545, Physica-B-319-122-2002, APL-89-113105-2006, APL-89-062512-2006} given by the lines in Fig.~\ref{fig:coercivity-deviation}.
The additional models, to be discussed later, are all based on a conventional micromagnetic approach. In such models the temperature dependence of magnetic properties is not intrinsic to the formalism, as it is in the atomistic approach, and must be introduced explicitly. 
Consequently, in the micromagnetic-based models we will introduce the effect of temperature ($T = 300$ K) by normalizing the micromagnetic parameters in the theoretical calculations. For the anisotropy constants, $K_\mathrm{core(edge)}$, we use the Callen-Callen law\cite{Callen19661271}
%==========================================
 \begin{equation}\label{eq:temperature-dependence-of-K}
K_\mathrm{core(edge)}(T=300 \ \textrm{K}) \approx K_\mathrm{core(edge)}(T=0 \; \mathrm{K})m_\mathrm{e}^{3}\mathrm{,}
\end{equation}
%==========================================
where $m_\mathrm{e}=M_\mathrm{s}(T=300 \ \textrm{K})/M_\mathrm{s}(T = 0 \ \textrm{K})=0.82$ is obtained directly from our computational atomistic spin calculations.
$M_\mathrm{s}$ is the saturation magnetization of the core (edge).
We note that experimentally the exponent value for the decrease in anisotropy constant of $L1_0$ FePt is close to 2.1,\cite{PhysRevB.66.024413} which can be reproduced using the multiscale atomistic spin model simulations.\cite{Mryasov-EPL-2005} However, this multiscale simulation is computationally expensive for the calculation of dots with the diameter of 25 nm [Fig.~\ref{fig:Simulated-dot}]. Therefore, we have used a simplified atomistic spin model, where $K_\mathrm{core(edge)}(T)$ follows the Callen-Callen law. In fact at room temperature the 2.1 scaling law and the Callen-Callen law give similar length scales, e.g. the exchange length or the domain-wall length, which can be estimated by the ratio $m_\mathrm{e}^{2.1/2}/m_\mathrm{e}^{3/2} \sim 0.93$. We also note that both surface and interface effects can slightly vary the Callen-Callen law for $K_\mathrm{edge}(T)$.\cite{PhysRevB.82.054415} 
The exchange stiffness of the core (edge), $A_\mathrm{core(edge)}(T)$, has been shown to scale with $m_\mathrm{e}$ as
\cite{Franz-GRL3841, PhysRevB.82.134440}
%==========================================
\begin{equation}
\label{eq:temperature-dependence-of-A}
	A_\mathrm{core(edge)}(T=300 \ \textrm{K}) \approx A_\mathrm{core(edge)}(T=0 \; \mathrm{K})m_\mathrm{e}^{1.745}\mathrm{.}
\end{equation}
%===========================================

\subsection{Narrow soft edge: individual nucleation}

The magnetization reversal in the absence of soft edges, as shown in the spin configuration snapshot for $W_\mathrm{edge} = 0$ nm in Fig.~\ref{fig:properties}(b),  starts by the nucleation of a small region (red area in the snapshot) in the boundary and proceeds with the subsequent expansion to the entire dot. At this point it is worthwhile considering the physical origin of the nucleated reversal. The origin of the incoherent nucleated reversal process lies in the combination of high magnetocrystalline anisotropy and thermal fluctuations. At applied fields in the vicinity of the coercivity thermal fluctuations break the symmetry of the dot and cause a nucleation event. The narrow domain wall width, arising from the high magnetocrystalline anisotropy in the core, stabilizes the nucleated domain. Following the nucleation the lowest energy barrier for switching is then propagation of the domain wall, leading to an incoherent reversal mechanism. The combination of short time scales, high anisotropy, and system size greater than the domain wall width gives the fundamental physical origin of the thermal switching field distribution. For longer timescales more nucleation attempts are made reducing the effective thermal SFD since the material switches at the same field, while for lower anisotropy materials the nucleated domain is unstable and so the effect of thermal fluctuations is also lower.

For dots with a narrow soft edge, $W_\mathrm{edge} = 1$ or $2$ nm, the reversal mechanism is the same as for dots with no soft edges, although due to the low coercivity of the edge the nucleation field is reduced significantly. The thermal SFD also reduces rapidly with narrow soft edges due to the reduced stability of the nucleated domain owing to the lower effective anisotropy. In an attempt to quantify the reduction in the coercivity as a function of the edge width we have developed an atomistic one-dimensional (1D) linear chain model, details of which are given in Appendix~\ref{sec:linear-chain}. By estimating the coercive field as an effective anisotropy field of the nucleated area, $H_\mathrm{K}^\mathrm{eff}$, the linear chain model predicts a linear decrease in the coercivity given by
%======================================
\begin{equation}
\label{eq:Hc-linear}
	H_\mathrm{c}^\mathrm{core(edge)} = H_\mathrm{K}^\mathrm{eff} = H_\mathrm{K}^\mathrm{core} \big( 1-bW_\mathrm{edge} \big)\mathrm{.}
\end{equation}
%======================================
In Fig.~\ref{fig:coercivity-deviation}(a) we can see that for $W_\mathrm{edge}\leq 2$ nm, both the coercivity of the core and the edge are equal and linearly decrease as a function of the edge width. It can be seen that Eq.~\ref{eq:Hc-linear} gives reasonable agreement with the numerical results.

\subsection{Wide soft edge: an incomplete to a complete circular domain wall}
With a further increase in the edge width we observe the reversed region with a negative curvature, shown by Fig.~\ref{fig:properties}(b) for $W_\mathrm{edge} = 3$ nm with nucleated areas denoted by red regions. The negative curvature could suggest that more than one reversed region nucleates during the reversal. The deviation between $H_\mathrm{core(edge)}$ and $H_\mathrm{K}^\mathrm{eff}$ [Eq.~\eqref{eq:Hc-linear}] reflects the reversal dominated by multi-reversed regions. These multiple nucleation events also mark an increasing difference between $H_\mathrm{c}^\mathrm{edge}$ and $H_\mathrm{c}^\mathrm{core}$ values with further increases in the edge width [Fig.~\ref{fig:coercivity-deviation}(a)]. From the spin configuration snapshots in Fig.~\ref{fig:properties}(b) we observe this behavior corresponds to an incomplete circular domain wall formed at the core/edge interface. In this region we cannot approach $H_\mathrm{core(edge)}$ using Eq.~\eqref{eq:Hc-linear} because this is only valid for the reversal dominated by a single reversed region. Instead, we find that $H_\mathrm{c}^\mathrm{edge}$ approaches the domain wall nucleation field $H_\mathrm{n}$, obtained from the analytical expression derived for the limit of strong hard/soft coupling with a soft layer thicker than the domain-wall width ($\approx 5$ nm in this study) in the hard layer\cite{goto:2951, PhysRevB.70.104405, Suess2009545} given by
%=======================================
\begin{equation}
\label{eq:nucleation-field}
	H_\mathrm{n}=H_\mathrm{K}^\mathrm{edge}+\Big(\frac{\pi}{2} \Big)^2\Big(\frac{l_\mathrm{EX}^\mathrm{edge}}{W_\mathrm{edge}}\Big)^2M_\mathrm{edge}\mathrm{,}
\end{equation}
%=======================================
 where $H_\mathrm{K}^\mathrm{edge}=2K_\mathrm{edge}/M_\mathrm{edge}$ is the anisotropy field of the edge, and $M_\mathrm{edge}$ is the saturation magnetization of the edge and $l_\mathrm{EX}^\mathrm{edge}=\sqrt{A_\mathrm{edge}/K_\mathrm{edge}}$ is the exchange length in the edge.

For applied fields larger than $H_\mathrm{n}$ but less than the domain-wall pinning field at the edge/core interface, $H_\mathrm{p}$, the increased field compresses the domain wall in the edge and therefore reduces the corresponding domain-wall width, $l_\mathrm{DW}^\mathrm{edge}$. At even wider edge widths [for example, see Fig.~\ref{fig:properties}(b) at $W_\mathrm{edge} = 7$ nm], the nucleation occurs in the entire edge, but the domain wall is then pinned at the core/edge interface, showing a circular domain wall. As the reversal continues, the domain wall propagates inwards until collapse and full magnetization reversal.
In addition, the propagated domain wall shows a non-circular symmetry [Fig.~\ref{fig:properties}(b) at $W_\mathrm{edge} = 7$ nm], in contrast to the circular symmetry of the domain wall pinned at the core/edge interface. The suggests the depinning of part of the circular domain wall during the reversal of spins in the core. However, in contrast to the analytical model of the single nucleation region proposed in Ref. 20, the reversed region in the core shows a negative curvature [Fig.~\ref{fig:properties}(b) at $W_\mathrm{edge} = 7$ nm], indicating that the reversal could be dominated by the multinucleation events.
 On the other hand, $H_\mathrm{c}^\mathrm{core}$ saturates at $H_\mathrm{p}$ when $W_\mathrm{edge} \geq l_\mathrm{DW}^\mathrm{edge}$, which reads\cite{Physica-B-319-122-2002, APL-89-113105-2006, APL-89-062512-2006}
 %=========================================
 \begin{equation}
 \label{eq:domain-wall-width-edge}
	l_\mathrm{DW}^\mathrm{edge}=\pi\sqrt{\frac{2A_\mathrm{edge}}{K_{\mathrm{core}}+K_{\mathrm{edge}}}}\mathrm{.}
\end{equation}
%=========================================
$H_\mathrm{p}$ is given by\cite{Physica-B-319-122-2002, APL-89-113105-2006, APL-89-062512-2006}
%===========================================
 \begin{equation}
 \label{eq:pinning-field}
	H_\mathrm{p}={\frac{1}{4}}{\frac{2\big[K_{\mathrm{core}}-K_{\mathrm{edge}}\big]}{M_{\mathrm{edge}}}}\mathrm{.}
\end{equation}
%===========================================
Figure~\ref{fig:coercivity-deviation}(a) shows that our simulation results fit perfectly to the $H_\mathrm{p}$ (black dashed line).
Thus it confirms that for soft edges wider than $l_\mathrm{DW}^\mathrm{edge}$ at $H_\mathrm{p}$, the reversal mechanism is through depinning of part of a circular domain wall at the edge/core interface driven by the multiple nucleation events in the core with $H_\mathrm{c}^\mathrm{core} = H_\mathrm{p}$.

\subsection{Thermally induced switching field distribution}
The calculated thermal switching field distribution $\sigma_\mathrm{core(edge)}$ from the simulations for different edge widths is shown in Fig.~\ref{fig:coercivity-deviation}(b). Similarly to the coercive fields, our simulations show that for $W_\mathrm{edge} \leq 2$ nm, $\sigma_\mathrm{core} \simeq \sigma_\mathrm{edge}$, the thermal SFD displays a linear decrease with the increasing $W_\mathrm{edge}$. In order to quantify the thermal fluctuations in the coercive field within a micromagnetic framework it is necessary to associate the magnetic moment $\mu$ in Eq.~\eqref{thermal-field} with a volume characteristic of magnetization reversal. For this we use the activation volume, $V_\mathrm{act}$, which is an equilibrium quantity and defined as the volume associated with the magnetization change between positions of minimum and maximum static energy.\cite{gaunt:4129} Furthermore, we average the thermal fluctuation field over a specific time equal to the inverse of an "attempt frequency" used in phenomenological models of thermal activation processes. The attempt frequency is generally taken as the natural frequency of oscillation in the local minimum, i.e., $f_0 = \gamma H_\mathrm{K}$ with $H_\mathrm{K}$ the anisotropy field. This leads to a variance in the field components, which we take as the standard deviation of coercivity, $\sigma_\mathrm{core(edge)}$, given by
%==============================================
\begin{equation}\label{thermal-SFD}
\sigma_\mathrm{core(edge)} = \sqrt{\frac{2 \lambda k_{\mathrm{B}} T H_\mathrm{K}}{M_\mathrm{s}V_\mathrm{act}}}\mathrm{.}
\end{equation}
%==============================================
For $W_\mathrm{edge} \leq 2$ nm, the single nucleated region dominates the reversal. However, the observed nucleation is a nonequilibrim quantity.\cite{PhysRevB.55.11521} For $V_\mathrm{act}$ one should estimate the volume of the equilibrium domain change during reversal. Considering the dot size is smaller than the domain size ($\sim 26$ nm in this study), we can treat the dot as a single domain particle and therefore approach $V_\mathrm{act}$ to the total volume of the dot, $V_\mathrm{act} \sim \pi(r_\mathrm{core}+W_\mathrm{edge})^2t_\mathrm{d}$. Taking $H_\mathrm{K} = H_\mathrm{K}^\mathrm{eff}$ [Eq.~\eqref{eq:Hc-linear}], we arrive at 
 %==============================================
\begin{equation}\label{thermal-SFD-narrow-edge}
\sigma_\mathrm{core(edge)} = \sigma_\mathrm{K} = \sqrt{\frac{2 \lambda k_{\mathrm{B}} T H_\mathrm{K}^\mathrm{eff}}{M_\mathrm{s}[\pi(r_\mathrm{core}+W_\mathrm{edge})^2t_\mathrm{d}]}}\mathrm{.}
\end{equation}
%============================================== 
where $\sigma_\mathrm{K}$ is $\sigma_\mathrm{core(edge)}$ in this region. It can be seen that Eq.~\eqref{thermal-SFD-narrow-edge} [indicated by the gray dashed line in Fig.~\ref{fig:coercivity-deviation}(b)] gives results reasonably close to the numerical results.

For $W_\mathrm{edge} \geq 3$ nm, the common behavior of spins in the core starts to deviate from that in the edge, as we show in Fig.~\ref{fig:coercivity-deviation}(a). Similarly we find that $\sigma_\mathrm{core}$ deviates from $\sigma_\mathrm{edge}$ [Fig.~\ref{fig:coercivity-deviation}(b)]. In this region, the different reversal mode of core spins with that of edge spins suggests that $V_\mathrm{act}$ in the edge approaches to the edge volume, $V_\mathrm{act} \sim \pi \big[ \big( r_\mathrm{core}+W_\mathrm{edge} \big) ^2-\big( r_\mathrm{core} \big) ^2 \big] t_\mathrm{d}$. Using Eq.~\eqref{thermal-SFD} with $H_\mathrm{K} \sim H_\mathrm{c}^\mathrm{edge} = H_\mathrm{n}$ [Eq.~\eqref{eq:nucleation-field}] $\sigma_\mathrm{edge}$ in this region, $\sigma_\mathrm{n}$, is [purple solid line in Fig.~\ref{fig:coercivity-deviation}(b)]
 %==============================================
\begin{equation}\label{thermal-SFD-nucleation}
\sigma_\mathrm{n} = \sqrt{\frac{2 \lambda k_{\mathrm{B}} T H_\mathrm{n}}{M_\mathrm{s}\pi \big[ \big( r_\mathrm{core}+W_\mathrm{edge} \big) ^2-\big( r_\mathrm{core} \big) ^2 \big] t_\mathrm{d}}}\mathrm{.}
\end{equation}
%==============================================
For $W_\mathrm{edge}\geq l_\mathrm{DW}^\mathrm{edge}$, $H_\mathrm{c}^\mathrm{core}$ saturates at $H_\mathrm{p}$ [Eq.~\eqref{eq:pinning-field}]. The different reversal behavior of core spins to that of edge spins bring us to the estimation of the activation volume in the core as the core volume, $V_\mathrm{act} \sim \pi \big[ \big( \big( r_\mathrm{core} \big) ^2 \big] t_\mathrm{d}$. Using Eq.~\eqref{thermal-SFD} with $H_\mathrm{K} \sim H_\mathrm{c}^\mathrm{core} = H_\mathrm{p}$ [Eq.~\eqref{eq:pinning-field}] we arrive at [black dashed line in Fig.~\ref{fig:coercivity-deviation}(b)]
 %==============================================
\begin{equation}\label{thermal-SFD-pinning}
\sigma_\mathrm{p} = \sqrt{\frac{2 \lambda k_{\mathrm{B}} T H_\mathrm{p}}{M_\mathrm{s}\pi r_\mathrm{core}^2 t_\mathrm{d}}}\mathrm{,}
\end{equation}
%==============================================
where $\sigma_\mathrm{p}$ is $\sigma_\mathrm{core}$ in this region. Eqations.~\eqref{thermal-SFD-nucleation} and~\eqref{thermal-SFD-pinning} gives values of $\sigma_\mathrm{core(edge)}$ roughly a factor of 2 different from the numerical results [Fig.~\ref{fig:coercivity-deviation}(b)]. Given the assumptions involved the difference is reasonable agreement.

\subsection{Effect of interfacial exchange coupling on the reversal modes}

%===================================================
\begin{figure}[!t]
\center
\includegraphics[width=8cm]{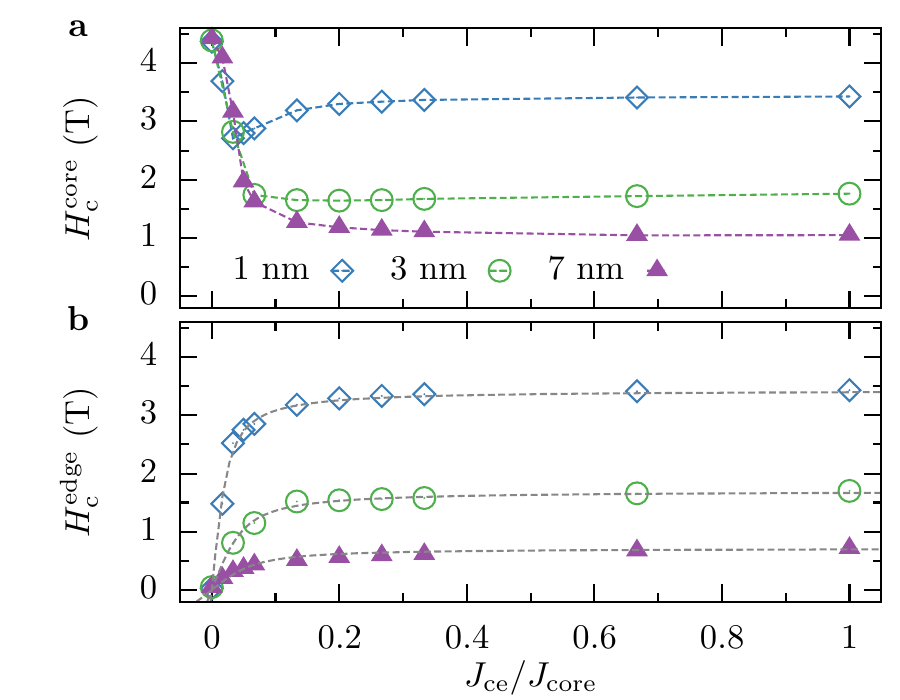}
\caption{(Color online) (a) Coercivity of the core as a function of the normalized core/edge exchange coupling strength at varied width of edges. The core/edge exchange coupling strength is normalized to the exchange interaction between spins in the core (edge). Dashed lines are guided by the eye. (b) Coercivity of the edge as a function of the normalized core/edge exchange coupling strength. Dashed curves represent a fitting to the Langevin function [Eq.~\eqref{eq:Hc-edge-coupling-strength}].}
\label{fig:Coercivity-edge-core-coupling}
\end{figure}

%===================================================

Finally we investigate the effect of core/edge exchange coupling strength, $J_\mathrm{ce}$, on the reversal modes in the nanodot. To do so we vary  the normalized interfacial exchange coupling strength $\tilde{J}_\mathrm{ce} = J_\mathrm{ce}/J_\mathrm{core(edge)}$ from 0 (no coupling) to 1 which corresponds to the strong coupling
studied in detail  in the previous  section. Here, we also perform hysteresis-loop calculations  for varied edge-width $W_{\textrm{edge}}$.

 Figure~\ref{fig:Coercivity-edge-core-coupling}(a) shows the coercivity of the core, $H_\mathrm{c}^\mathrm{core}$, as a function of $\tilde{J}_\mathrm{ce}$  for $W_{\textrm{edge}}=1,3$  and $7$ nm to cover  the three well-separated regimes of reversal modes observed in our system.
We observe that for $W_{\textrm{edge}}=1$ nm the core coercive field, $H_\mathrm{c}^\mathrm{core}$, presents a minimum at  a relatively weak coupling strength, similar to results observed in hard/soft structures, where the observed minimum is related to the two-spin behavior.\cite{richter:08Q905} Considering that the local minimum of $H_\mathrm{c}^\mathrm{core}$ happens when $J_\mathrm{edge} \gg J_\mathrm{ce}$ ($\tilde{J}_\mathrm{ce} \leq 1/20$) in narrowed edges ($W_\mathrm{edge} \leq 2$ nm), all spins in the edge might behave as a single macrospin. During the reversal, the single-spin behavior in the edge could provide a torque to spins in the core and yield the local minimum of $H_\mathrm{c}^\mathrm{core}$, which has been previously observed in the two-spin model\cite{richter:08Q905} as well as in the experiment.\cite{wang:142504}

As the coupling increases, the coercive field saturates to some value which has been already discussed in previous sections of the present work.
For edge widths larger than or equal to 3 nm, the minimum of the core coercive field disappears and a monotonous decrease in $H_\mathrm{c}^\mathrm{core}$ to a saturation value is observed.
For $W_{\textrm{edge}} \geq 7$ nm [equal to $l_\mathrm{DW}^\mathrm{edge}$ given by Eq.~\eqref{eq:domain-wall-width-edge}], this saturation value corresponds to the domain-wall depinning field.
Therefore, the interface coupling dependence of the core coercive field for $W_{\textrm{edge}} \geq 7$ nm is similar.

On the other hand, the edge coercive field, $H_\mathrm{c}^\mathrm{edge}$, consistently  increases with increasing interfacial exchange coupling to a saturation value [see Fig.~\ref{fig:Coercivity-edge-core-coupling}(b)] following a
Langevin law representing the effective bias field created in the edge by the coupling to the core,  in direct analogy to a paramagnet magnetization in the presence of an external field and  thermal fluctuations.\cite{Langevin1905}
This effective bias field is comparable to the external field applied in the calculation of hysteresis loops and can be estimated by

 %=========================================
 \begin{equation}
 \label{eq:Hc-edge-coupling-strength}
	H_\mathrm{c}^\mathrm{edge}(\tilde{J}_\mathrm{ce}) = H_{\mathrm{c},1}^\mathrm{edge}L \big(
	\beta \mu_{\mathrm{edge}}H_\mathrm{ex}
	 \big)\mathrm{.}
\end{equation}
%===========================================
where $H_\mathrm{ex}$ estimates the average  effective  bias field in the edge  induced by the interfacial coupling, and $\mu_\mathrm{edge}=M_\mathrm{edge}V_\mathrm{edge}$ is the saturation magnetization of the edge.
The Langevin function is $L(x)=\coth(x)-1/x$.
$H_{\mathrm{c},1}^\mathrm{edge}$ is a fitting constant and coincides with $H_\mathrm{c}^\mathrm{edge}$ at $\tilde{J}_\mathrm{ce} = 1.0$ (calculated in the previous sections).
We  can assume that $\mu_{\mathrm{edge}}H_\mathrm{ex}=V_\mathrm{edge} d \tilde{J}_\mathrm{ce}$ where $d$ is a parameter that measures the energy transferred from the core to the edge via the interfacial coupling.
This parameter is expected to depend on the volume of the edge, $V_{\textrm{edge}}\sim W_{\textrm{edge}}^2 t_{\mathrm{d}}$,
so that as the thickness is fixed for all $W_\mathrm{edge}$, we expect that $d \sim 1/W_\mathrm{edge}^2$ similar to that in a soft/hard bilayer structure, $H_\mathrm{ex}\propto 1/t_\mathrm{soft}^{2}$.\cite{goto:2951}
In  Fig.~\ref{fig:Coercivity-edge-core-coupling}(b) we show that indeed this relation fits very well to simulations.

\section{Discussion and Conclusions}
To summarize, using atomistic spin model simulations, we have investigated reversal modes in patterned $L1_{0}$ FePt dots with damaged edges in the presence of thermal fluctuations. Specifically, the calculated dot is composed of a hard magnetic core, which represents the undamaged part of the dot, and the damaged edge with soft magnetic properties. We have investigated the effects of the extent of damage on the edge by varying its width. We observe that the nucleation initiates reversal for all width of edges. The increased edge width linearly decreases and then saturates the required field for nucleation, with the curvature of the initially nucleated region reducing from positive to negative. Furthermore, the increased edge width reduces the thermally induced switching field distribution, which is found related to both the nucleation field and the activation volume. We have further studied reversal modes in dots with varied core/edge interfacial coupling strength, which could possibly result from the core/edge interfacial roughness. For dots with narrow edges, the reversal behaves in a similar way with that obtained in the two-spin model, suggesting that we can treat all spins in the edge as a single effective macrospin. In addition, we describe the coercivity of the edge using the Langevin function, representing the competition between the effective field generated from the core/edge coupling strength and the thermal fluctuations.

While the numerical simulation by the atomistic spin model is sufficient to explain the magnetization reversal, it is insightful to digest these results by simpler analytic methods so that the key physics can be highlighted. In some cases studied here, the reversal dynamics of the minority spins is mainly one dimensional. We are thus motivated to employ the linear spin chain model to capture the one-dimensional dynamics. Specifically at narrow edges the linear chain model is able to estimate the required field for the nucleation. As the edge width increases the nucleation field of core spins fits to the domain-wall pinning field at the core/edge interface. Considering the computationally intensive nature of the atomistic spin model simulation, these analytic theories can provide a global sketch for different parameters at minimal costs.

Comparing to previous studies focused on the reversal modes along the layer-growth direction in the typical exchange spring media, here we present detailed two-dimensional reversal behaviors on the patterned dot-plane as well as the corresponding thermally induced switching field distribution, both of which in fact dominate properties of typical patterned dots and cannot be investigated by standard micromagnetic calculations. We also note that different magnetic properties of the edge, which have been assumed constant values in this study and have not been experimentally probed, only vary the characteristic length of different reversal modes and the corresponding coercivity fields without affecting the validity of theories. According to our study here, the presence of damaged edges with uniform magnetic properties reduces the thermally induced switching field distribution, and the width of the damaged edge significantly changes the coercivity in patterned dots. Therefore, the experimentally observed broadening of the switching field distribution in patterned $L1_0$ FePt dots with damaged edges\cite{IEEE-49-693-2013} should be attributed to extrinsic properties of the nanodots created by patterning processes, for example, the variation in either the width or the magnetic properties of the damaged edges.
As long as we can precisely control properties of damaged edges by applying a proper patterning technique, for example, ion implantation,\cite{PhysRevB-85-214417-2012} we could realize exchange spring bit-patterned media without additional soft layers.

\begin{acknowledgements}
The authors would like to thank  H.-H. Lin for insightful suggestions. Fruitful discussions with O. Hovorka,  J. Wu, W. Fan, P. Chureemart, and S. Ruta are also acknowledged. J.-W. also highly appreciates the assistance from J. Barker and  T. Ostler on solving computational issues. This work has been supported by the Ministry of Science and Technology under Grant No. MOST 101-2917-I-007-016. U.A. gratefully acknowledges support from Basque Country Government under "Programa Posdoctoral de perfeccionamiento de doctores del DEUI del Gobierno Vasco". The financial support of the Advanced Storage Technology Consortium (ASTC) and EU Seventh Framework Programme under Grant Agreement No. 281043 FEMTOSPIN is also gratefully acknowledged.
\end{acknowledgements}

\appendix

\section{Linear chain model}\label{sec:linear-chain}

%===================================================
\begin{figure}[!t]
\center
\includegraphics[width=8cm, trim=0 0 0 0]{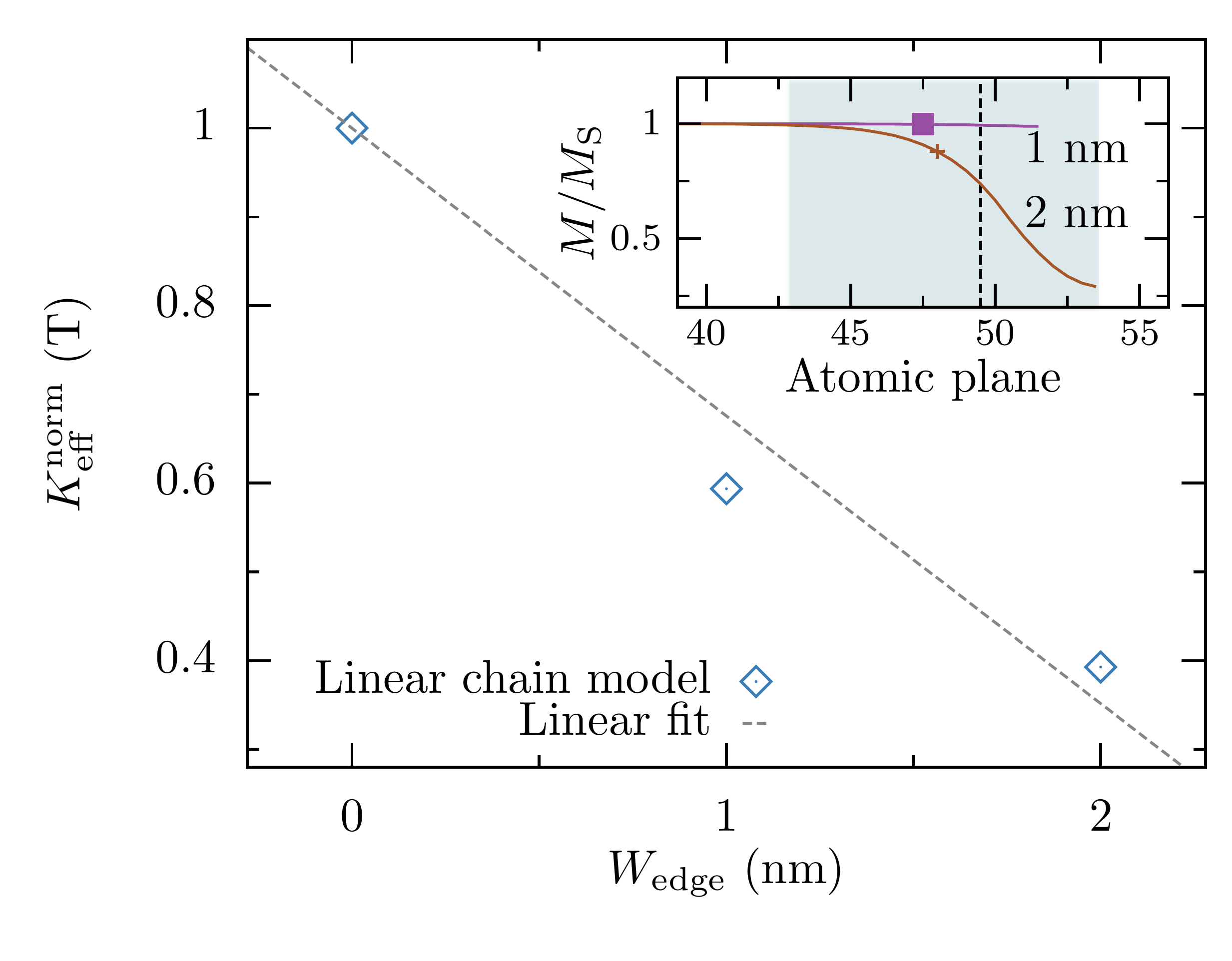}
\caption{(Color online)  Effective value of anisotropy, calculated by the linear chain model, as a function of the width of the edge. The gray dashed line is the linear fitted function. Inset shows the calculated layer-resolved magnetization prior to magnetization reversal in the core in the linear chain model. The atomic plane is counted from the center of the core toward the edge, and the vertical dashed line denotes the core/edge interface. The blue area indicates the nucleated region for the edge-width of 2 nm.}
\label{fig:effective-Ku}
\end{figure}
%====================================================

In order to quantify the variation of the coercivity for narrow edge thicknesses we have developed a 1-D atomistic linear chain model, simplifying the whole dot into a one-dimensional region started from the center of the core to the edge. Each spin in the chain model represents the average spin within a given atomic plane, and we can write down the following spin Hamiltonian
%=========================================
\begin{equation}
\label{eqn:1-D-Hamiltonian}
	\mathcal{H}_{i'} =  -\sum_{i', j'} J_{i'} \bsh_{i'} \cdot \bsh_{j'} - k_{i'} \left(\bsh_{i'}^z\right)^2 - \mu_{i'} \bHapp \cdot \bsh_{i'}\mathrm{,}
\end{equation}
%=========================================
where $i'$, $j'$ label different spins with the identical moment $\mu$. $J$ is the intra-layer exchange coupling, $\bsh$ is the unit vector representing the spin direction, $k$ is the anisotropy constant, and $\bHapp$ is the external applied field.
We set $\mu = \mu_\mathrm{core} = \mu_\mathrm{edge} = 1.5$ $\mu_\mathrm{B}$, $k = k_\mathrm{core} = 4.9 \times 10^{-23}$ J/link for spins in the core and  $k = k_\mathrm{edge} = 1 \times 10^{-24}$ J/link for those in the edge.
We allow reduced exchange coupling at the core/edge interface by writing the exchange energy between interface spins as
%=========================================
\begin{equation}
\label{eqn:Interface-Hamiltonian}
	\mathcal{H}_\mathrm{int} =  J_\mathrm{int}\bsh_{i'} \cdot \bsh_{\nu '} \mathrm{,}
\end{equation}
%=========================================
where $J_\mathrm{int}$ is the interface exchange coupling, and $\nu '$ labels spins in separate regions (core or edge) from those labeled by $i'$.

The equilibrium state of the spin system is determined by solving the Landau-Lifshitz equation, with no precession term
%=========================================
\begin{eqnarray}
\label{eqn:LL}
	\frac{\partial\bsh_{i'}}{\partial t} = -\frac{\gamma}{(1+\lambda^{2})} \bsh_{i'} \times \lambda (\bsh_{i'} \times \bold{H}_{{i'}, \mathrm{eff}})\mathrm{,}
\end{eqnarray}
%=========================================
where $\lambda$ is the intrinsic damping parameter, $\gamma$ is the absolute value of the gyromagnetic ratio, and $\bold{H}_{{i'}, \mathrm{eff}}$ is the effective magnetic field in each atomic plane, given by
%=========================================
\begin{equation}\label{eqn:Heff-1D}
  \bold{H}_{{i'}, \mathrm{eff}} = -\frac{1}{\mu_{i'}}\frac{\partial \big(\mathcal{H}_{i'} + \mathcal{H}_\mathrm{int} \big)}{\partial \bold{S}_{i'}} \mathrm{.}
\end{equation}
%=========================================
In the inset of Fig.~\ref{fig:effective-Ku}, we show the calculated layer-resolved magnetization within the spin chain model with various $W_\mathrm{edge}$, after positively saturating all spins and then applying a corresponding negative field prior to magnetization reversal in the core.
We number the atomic plane from the center of the core to the edge, and the vertical dashed line in the inset of Fig.~\ref{fig:effective-Ku} denotes the core/edge interface. Increasing $W_\mathrm{edge}$ gives rise to increasing penetration of the domain wall into the core. From the energy contributed to the reversal, we estimate a normalized effective value of the anisotropy constant, $K_\mathrm{eff}^\mathrm{norm}$, by integrating the anisotropy energy over the domain-wall width from the edge to the core in the nucleated region (see the blue region in the inset of Fig.~\ref{fig:effective-Ku} for $W_\mathrm{edge} = 2$ nm) and then normalizing to $K_\mathrm{core}$. This quantifies the reduction in the energy barrier due to the exchange spring. Figure~\ref{fig:effective-Ku} illustrates the variation of $K_\mathrm{eff}^\mathrm{norm}$ with $W_\mathrm{edge}$. We observe a linear decrease in $K_\mathrm{eff}^\mathrm{norm}$ with the increase in $W_\mathrm{edge}$, and we further describe the linear decrease as (gray dashed line in Fig.~\ref{fig:effective-Ku})
%=====================================
\begin{equation}
\label{eq:effective-Ku}
	K_\mathrm{eff}^\mathrm{norm}=1-bW_\mathrm{edge}\mathrm{,}
\end{equation}
%========================================
where $b = 0.324$ (nm$^{-1}$) obtained from fitting.
Since a single nucleated area dominates the reversal in the region of narrow soft edges, we then estimate the coercive field  as an effective anisotropy field of the nucleated area, $H_\mathrm{K}^\mathrm{eff}$, indicated by the gray dashed line in Fig.~\ref{fig:coercivity-deviation}(a),
%======================================
\begin{equation}
\label{eq:Hc-linear2}
	H_\mathrm{c}^\mathrm{core(edge)} = H_\mathrm{K}^\mathrm{eff} = H_\mathrm{K}^\mathrm{core} \big( 1-bW_\mathrm{edge} \big)\mathrm{.}
\end{equation}


%\bibliography{reference-3}
\end{document}